\documentclass[showpacs,prb,letterpaper,floatfix,nobalancelastpage,twocolumn]{revtex4}
\usepackage{dcolumn}
\usepackage{bm}
\usepackage{amsmath}
\usepackage{bm,braket}
\usepackage{color}
\usepackage{float}
\usepackage{graphicx,subfigure}
\usepackage{amssymb}
\usepackage{sidecap}
\usepackage{float}

\newcommand{\bfr}{\bf r}
\newcommand{\bfg}{\bf G}
\newcommand{\bfd}{\bf d}

\begin{document}




\title{Quantum dynamics of two quantum dots coupled through localized plasmons:\\ An  intuitive and accurate quantum optics approach using quasinormal modes}

\author{Rong-Chun Ge}
\email{rchge@physics.queensu.ca}
\author{Stephen Hughes}
\email{shughes@physics.queensu.ca}
\affiliation{
Department of Physics, Engineering Physics and Astronomy, Queens University, Kingston, Ontario, Canada K7L 3N6}
\newcommand{\com}[1]{{\color{blue}#1}}
\begin{abstract}
We study the quantum dynamics of two quantum dots (QDs) or artificial atoms  coupled through the fundamental localized plasmon of  a gold nanorod resonator. We  derive  an intuitive and efficient time-local master equation, in which the effect of the metal nanorod is taken into consideration self-consistently using a  quasinormal mode (QNM) expansion technique of the photon Green function. Our efficient QNM technique offers an alternative and more powerful approach over the standard Jaynes-Cummings model, where the radiative decay, nonradiative decay, and spectral reshaping effect of the electromagnetic environment is rigorously included in a clear and transparent way. We also show how one can use our approach to compliment the approximate Jaynes-Cummings model  in certain spatial regimes where it is deemed to be valid. We then present a study of the quantum dynamics and photoluminescence spectra of the two plasmon-coupled QDs.  We first explore the non-Markovian regime, which is found to be important only on the ultrashort time scale of the plasmon mode which is about 40$\,$fs. For the field free evolution case of excited QDs near the nanorod, we demonstrate how spatially separated QDs can be effectively coupled through the plasmon resonance and we show how  frequencies away from the plasmon resonance can be more effective for coherently coupling the QDs. Despite  the strong inherent dissipation of gold nanoresonators, we show that qubit entanglements as large as 0.7 can be achieved from an initially separate state, which has been limited to less than 0.5 in previous work for weakly coupled reservoirs. We also study the superradiance  and subradiance decay dynamics of the QD
pair. Finally, we investigate the rich quantum dynamics of QDs that are incoherently pumped,
and  study the  polarization dependent behaviour of the emitted photoluminescence spectrum where a double-resonance structure is observed due to the strong photon exchange interactions. Our general quantum plasmonics formalism can easily be extended to include multiple  QDs  interacting through the QNMs of metallic resonator structures, fully accounting for radiative and non-radiative coupling, as well as nonlinear  light-matter interaction processes.

\end{abstract}

\pacs{42.50.Nn, 73.20.Mf, 78.67.Bf}

\maketitle
\section{introduction}

Quantum nanophotonics is an active field of research, which is driven in part by fundamental ideas in light-plasmon-matter interactions, applications in nanophotonics, and  by continued advances in nanofabrication technologies. In particular, various types of nanoscale  structures have been designed and fabricated to manipulate the interaction between quantum emitters and local electric fields, which can be enhanced by  tailoring the local density of optical states (LDOS) typically through some discrete cavity resonance~\cite{Vahala,TameQP}. For coupling to electric fields below the diffraction limit,  metallic nanoparticles (MNPs) have been shown to yield an unprecedented confinement of light at the nm scale due to the spatial structure of the localized surface plasmon (LSP) resonances. For quantum dot (QD) emitters  or artificial atoms placed  sufficiently close to the MNP,  the strong coupling regime has also been shown to be experimentally accessible~\cite{colestrongcoupling,Hohenester:PRB08,strongnanoshell,FGGVidal}. As a consequence of the extreme spatial confinement of the LSP, the corresponding effective mode volume of the electric field is much smaller than more traditional dielectric cavity structures, which leads to a strong enhancement of the spontaneous emission (SE) rate in the weak-to-intermediate coupling regime~\cite{Novotny06,Frimmer13,Belacel}.
Moreover, fine spatial control of the QD dynamics at the single quantum excitation level and processing of the light signal at the nanoscale is possible~\cite{TameQP,KrPRL109,Lee14}, resulting in a broad range of applications in fields such as high precision quantum information processing and quantum computation, efficient solar cells~\cite{opticalantenna}, and high precision chemical or biological detection~\cite{KR07,KYHLIRM97,Hou13}.

Although long distance and large scale transmission of information using  metallic structures is typically not practical because of the strong inherent Ohmic losses of metals  at optical frequencies, improvements
can be made by  using hybrid semiconductor-metallic structures~\cite{Oulton08,Sorger11}, in which the transmission is carried out by semiconductor optical technology, while a nanoscale metallic resonator can be used as an effective modulator~\cite{KrPRL109} and/or transistor~\cite{Chang07}. Thus it is of significant 
fundamental and applied interest to  study the interaction between dipole emitters such as QDs and individual MNPs. Recently,  there have been a few works  studying  how quantum emitters couple to  LSPs,
e.g., to describe enhanced  SE (Purcell effect)~\cite{Novotny06,Frimmer13,Belacel}, entanglement dynamics~\cite{plasmonentanglement}, and the fluorescence spectrum~\cite{ourprb13}. 
In the classical or semiclassical regime, with the exception of a particular type of simple geometry 
such as a spherical cavity~\cite{aplsun}, for which classical analytical results are available, most of the nanoplasmonic studies are carried out by numerical analysis
which is numerically cumbersome and not  physically intuitive~\cite{FDTD00,Busch1,FEM09}.
In the quantum optics regime of cavity-QED (cQED), it has been common to exploit a standard cavitylike master equation with phenomenological decay rates that are implicitly Lorentzian in their decay dynamics~\cite{Waks10,Savasta:PRL2010,GrayPRB13}; such an approach is useful and easy to understand, but it is ultimately limited since the general non-Lorentzian nature of the LSP is neglected,
and it it not clear  how to obtain the various coupling parameters, e.g., as a function of QD distance from the MNP resonator.

As a result of the fluctuation-dissipation theorem, in general a continuous mode theory instead of the simple single mode theory needs to be employed for a quantum optics description of an inhomogeneous lossy structure, and a quantum noise term can be included phenomenologically~\cite{TPRA51,StefanPRA58} or by including a continuous reservoir at the level of a microscopic theory~\cite{BrunoPRA46,LPRA70}; both of these approaches result in a powerful framework with the continuous response of the medium embedded in the medium Green function, which is obtained from an electric dipole source in Maxwell's equations.
On the other hand, it is highly desirable to be able to describe the physics of LSPs in terms of one or a few discrete modes, which has been the standard approach in dielectric cQED systems.
Recently, it is shown that the LSP can be effectively described as the
quasinormal modes (QNMs) of the MNP~\cite{quasi3,PhilipACS}, which are defined as the eigenfunctions of wave equation in the frequency domain with open boundary conditions~\cite{Leung941,Lee99}. A generalized mode expansion technique of the classical photon Green function based on QNMs has been shown to work extremely well for various shaped MNPs, and the SE enhancement of an electric dipole located both inside and outside of MNP shows excellent agreement with  full numerical calculations over a broad range of frequencies around the the LSP resonance~\cite{quasius1,quasius2,ourlocal}.
The combination of an insightful QNM approach and a rigorous Green function approach to quantum optics is  thus highly desired, as MNPs facilitate a coupling regime, in general with a non-Lorentzian spectral density, i.e., beyond a dissipative Jaynes-Cummings (JC) model. In certain limits, it can also be used to aid a JC model and justify when such a simpler model can work, with clearly identified coupling rates that can be obtained from QNM theory. Indeed, the MNP  yields a rich mode coupling regime as a function of position and polarization, and allows one to explore a complex interplay of radiative and nonradiative dynamics that are unique to the metal environment.

\begin{figure}
\centering
\includegraphics[trim=0.0cm 5.0cm 1.3cm 0.0cm, clip=ture, width=.99\columnwidth]{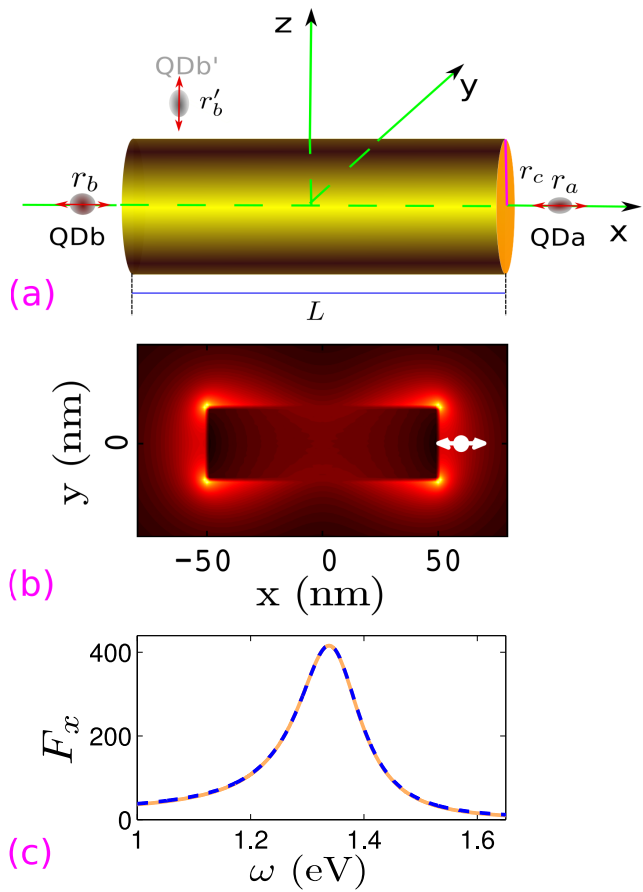}
\caption{(Color online) (a) Schematic  of the QD nanorod system with background refractive index, $n_B = 1.5$;  QD positions (the dark brown/light grey ellipsoids) are indicated near both ends of the gold nanorod; the nanorod  has the dimensions $L=100$~nm and $r_{ c}=15~$nm. (b) Norm of the QNM, $|\tilde{\bf f}_{\rm c}(x,y,z=0)|$, with complex eigenfrequency $\tilde{\omega}_{\rm c}/2\pi = (\omega_{\rm c}+{\rm i}\gamma_{\rm c})/2\pi = 324.981 - {\rm i}16.58~$THz. Yellow (bright) color indicates the highest intensity regions. (c) Enhancement of the $x$-projected LDOS/SE (see text), $F_x$ [Eq.~(\ref{enhancement})], for an $x$-polarized QD, at ${\bf r}_a = (60, 0, 0)$~nm---as is shown by QD$a$/white arrow in (a)/(b); the orange (solid)/blue (dashed) lines are given by Eq.~(\ref{qme}) and full-dipole numerical calculations, respectively.
}
\label{fig:Sf1}
\end{figure}

In this paper we present a quantum optics  framework to model the quantum dynamics between two QDs coupled to the LSP of a MNP system. The extension to model more than two QDs is straightforward and also described. While there  have been several papers studying the dynamics of two QDs coupled by a MNP~\cite{twodots1,twodots2}, these approaches, similar to the methods mentioned above, start from assuming the system could be described by the standard cQED master equation by adding in phenomenological decay parameters  by hand; by doing so, they neglect the possible non-Lorentzian features of the LDOS which is important in the case  for QDs that are sufficiently close to the LSP resonator~\cite{ourprb13}, and they do not incorporate the full
electromagnetic response of the MNP environment, including both radiative and nonradiative coupling effects. Instead of assuming a standard Lorentzian decay rate of the LSP, we start from a microscopic model and derive a master equation that takes into consideration the electromagnetic response of the MNP in detail by exploiting a  QNM expansion technique for the photon Green function~\cite{quasius1,quasius2}. As an example application of this theory, we consider two QDs in the vicinity of a gold
nanorod, as shown in Fig.~\ref{fig:Sf1}(a). While other MNP shapes can also be used in our theory, including metal dimers~\cite{quasius2}, the single nanorod is partly motivated  by the following reasons:
 ($i$) the LSP resonance is around 1.4 eV, which is close to the wavelengths used in optical communication and for many QD emitters, ($ii$) the nanorod is a nontrivial geometry for which analytic methods are not readily available, and $(iii),$ it is dominated
by a single cavity mode, polarized along the axis of the rod. While the technique  we exemplify below is a single mode theory,  it can  easily be generalized to include multiple LSP modes if there are several QNMs in the frequency regime of interest, and it properly includes the QNM dissipation. 
Recently, Yang {\em et al.} have carried out a somewhat similar effort, to study
the  simple  linear optical properties of a single dipole coupled to a metal
resonator \cite{Yang2015}.

The layout of our paper is as follows. In Sec.~\ref{sec:theory}, we present our main theoretical technique and derive a quantum master equation based on a rigorous quantum optics approach for the medium in terms of the photon Green function, which is obtained from the QNM of the LSP. In Sec.~\ref{sec:QNMJC}, we compare the QNM technique to the JC model, and present the  improvements over the standard JC model with the help of the QNM technique. We also discuss how our approach could be used in conjunction with the driven JC model in certain regimes, providing a rigorous definition for the various coupling parameters.  In Sec.~\ref{sec:resulta},
our first example studies the simple SE dynamics from a single QD located around the metal nanorod, and shows that the non-Markovian dynamics is important on a time scale of around the lifetime of the LSP. In Sec.~\ref{sec:resultsb}, we study the free-field dynamics of two QDs in a homogeneous background, coupled by the nanorod LSP and show that  two qubit (QD) entanglement can be established within a few picoseconds for separate states with a peak value  larger than 0.7, despite the strong  Ohmic losses; we also study the affect of QD pure dephasing  on the peak value of the entanglement evolution, and investigate the concurrence (as a measure of entanglement) for different QD  distances from both sides of the nanorod. In Sec.~\ref{sec:resultsc}, we  study the incoherent spectrum for the excited two-QD system; in particular, we show explicitly how the real and imaginary part of the Green function contributes to the coupling between spatially separated QDs, and find a rich polarization-dependent behavior of the spectra,  including a double-resonance feature which is mediated by the strong photon exchange effects.   We present our conclusions in Sec.~\ref{sec:con}.

\section{Theory: quantization scheme for a general medium, master equation, and Green function expansion in terms of quasinormal modes}
\label{sec:theory}
For our MNP we consider a 3D gold nanorod as shown in Fig.~\ref{fig:Sf1}(a) with two QDs (dark brown/light grey ellipsoids), QD $a$ and $b$, located around both ends of the nanorod.  We  use parameters for a metal nanorod with length $L=100~$nm and radius $r_{ c} = 15~$nm, with the Drude model for the dielectric constant, $\varepsilon(\omega) = 1-\omega_{\rm p}^2/(\omega^2+{\rm i}\omega\gamma)$, where $\omega_{\rm p} = 1.26\times 10^{16}~$rad/s (bulk plasmon frequency) and $\gamma = 1.41\times 10^{14}~$rad/s (collision rate),  similar to the parameters for gold. The background refractive index is $n_B = 1.5$. We will also allow for the possibility of an incoherent pump field on the QDs. 

For the medium quantization scheme, we start from the formalism developed by 
Scheel/Dung {\em et al.}~\cite{Dung98,StefanPRA58,WelschPRA99}, which has been widely used to describe the quantum electrodynamics of a quantum emitter around a spherical metallic nanoresonator~\cite{colestrongcoupling,ourprb13,coherent}. This approach, however, is completely general and can be employed for any lossy inhomogeneous structure as long as the corresponding Green function ${\bf G}({\bf r},{\bf r}';\omega)$ can be calculated.
 The photonic Green function is defined through $\nabla \times \nabla \times {\bf G}({\bf r},{\bf r}';\omega) -k_{0}^{2} \varepsilon({\bf r},\omega) {\bf G}({\bf r},{\bf r}';\omega) = k_{0}^2\delta({\bf r}-{\bf r}'){\bf 1}$,
 for the position dependent complex permittivity $\varepsilon({\bf r},\omega)=\varepsilon_\text{R}({\bf r},\omega)+{\rm i}\varepsilon_\text{I}({\bf r},\omega)$, where $\varepsilon({\bf r},\omega)=\varepsilon(\omega)$ inside the nanorod and $\varepsilon({\bf r},\omega)=\varepsilon_B=n_B^2$ elsewhere; here
$k_0=\omega/c$ 
and ${\bf 1}$ is the unit dyadic. The imaginary part of the Green function with the same position arguments, ${\bfg}({\bfr},{\bfr};\omega)$ gives the
projected  LDOS $\propto {\rm Im}[{\bfg}_{ii}({\bfr},{\bfr};\omega)]/\omega$~\cite{nanoptics,Greenfunction}, while the Green function with different position arguments gives the propagator of the electric field. For a homogeneous dielectric,  the imaginary part of the homogeneous Green function is given by ${\rm Im}[{\bf G}_B({\bf r},{\bf r};\omega)]=\frac{\omega^3n_B}{6\pi c^3}{\bf 1}$.  

By treating the QDs as  two-level systems, and using the dipole and rotating-wave approximations, the total QD-MNP system is  described by the  Hamiltonian,
\begin{align}
H& = \hbar \int d{\bf r} \int_0^\infty d\omega  \, \omega \,{\bf f}^\dagger({\bf r},\omega) \cdot{\bf f}({\bf r},\omega) +\sum_{n=a,b}\hbar\omega_n\sigma_n^+\sigma_n^-
\nonumber \\
&- \bigg[\sum_{n=a,b}\sigma_n^+ \int_0^\infty \! d\omega\, {\bf d}_n \cdot {\bf E}({\bf r}_n,\omega) + {\rm H.c.}\bigg],
\end{align}
where $\sigma_n^+/\sigma_n^-$ (with $\sigma_n^{11} = \sigma_n^+\sigma_n^-$) are the Pauli matrices of the two QDs excitons (electron-hole pairs), and $\omega_n$ and ${\bf d}_n$ are the resonance frequency and dipole moment of the $n$-th QD, respectively;  ${\bf f/f^{\dagger}}$ are the boson field operators~\cite{WelschPRA99}, where the electric field operator is given by ${\bf E}({\bf r},\omega)={\rm i}\int d{\bf r}'{\bf G}({\bf r},{\bf r}';\omega)\cdot\sqrt{\frac{\hbar\varepsilon_I({\bf r}',\omega)}{\varepsilon_0\pi}}{\bf f}({\bf r'},\omega)$, with $\varepsilon_I({\bf r},\omega)$ the imaginary part of $\varepsilon({\bf r},\omega)$.

In a rotating frame at the frequency of the QD$a$, $\omega_a$, the total Hamiltonian becomes $H=H_{\rm S}+H_{\rm R} + H_{\rm I}$, where the system, reservoir, and the interaction terms are respectively defined by
\begin{subequations}
\begin{align}
H_{\rm S} = &-\hbar\Delta_{ab}\sigma^+_b\sigma^-_b,\\
H_{\rm I}=&-\sum_{n=a,b}\left(\sigma^+_n e^{i\omega_at}\int_0^{\infty}d\omega\,{\bf d}_n\cdot{\bf E}({\bf r}_n,\omega)+{\rm H.c.}\right),\\
H_{\rm R}=&\hbar\int dr\int_0^{\infty}d\omega\,\omega {\bf f}^{\dagger}({\bf r},\omega)\cdot{\bf f}({\bf r},\omega),
\end{align}
\end{subequations}
where $\Delta_{nm} = \omega_n-\omega_m$.
Transforming into the interaction picture, and using the second-order Born-Markov approximation, the  master equation for the reduced operator for QD pair is obtained from
\begin{align}
\frac{\partial}{\partial t}\tilde{\rho}(t) = -\frac{1}{\hbar^2}\int_0^{t}d\tau\,{\rm Tr}_{\rm R}\left\{[\tilde{H}_{\rm I}(t),\,\,[\tilde{H}_{\rm I}(t-\tau),\,\,\tilde{\rho}(t)\rho_R]]\right\},
\label{MH}
\end{align}
where $\rho_R=\rho_{R}(0)$ is the state of the reservoir; here we have assumed a second-order Born approximation, which is valid in the weak-to-intermediate coupling regime.
 We assume the temperature of the reservoir is $0$~K, which is a good approximation at optical frequencies. The field operators satisfy the following relations: ${\rm Tr}_R[{\bf f}^{\dagger}_i({\bf r},\omega),{\bf f}_j({\bf r}',\omega')\rho_R]=0$, ${\rm Tr}_R[{\bf f}_i({\bf r},\omega),{\bf f}^{\dagger}_j({\bf r}',\omega')\rho_R]=\delta_{ij}\delta({\bf r}-{\bf r}')\delta(\omega-\omega')$. After calculating the integrand on the right hand side of Eq.~(\ref{MH}) explicitly, we transform back to the Schr\"{o}dinger picture, and the generalized master equation for the reduced system is obtained as 
\begin{align}
\frac{\partial \rho}{\partial t}=\frac{1}{i\hbar}[H_{\rm S},\,\rho]+\int_0^td\tau \Big(\,\sum_{n,m}[\sigma_n^-(-\tau)\rho\sigma_m^+ -&\sigma_n^+\sigma_m^-(-\tau)\rho ]\nonumber\\
\times J_{\rm ph}^{nm}(\tau)+{\rm H.c.}\Big)+\sum_n\Big(\frac{\gamma_n^{\rm '}}{2}L[\sigma_n^{11}]+\frac{P_{n}}{2}L&[\sigma_n^+]\Big).
\label{ms}
\end{align}
Here   $J_{\rm ph}^{nm}(\tau)=\int_0^{\infty}d\omega 
J_{\rm ph}^{n m}(\omega)e^{i\tau(\omega_a-\omega)}$, with the photon reservoir function defined through 
\begin{equation}
J_{\rm ph}^{n m}(\omega)=\frac{{\bf d}_n\cdot{\rm Im}[{\bf G}_{nm}(\omega)]\cdot{\bf d}_m}{\pi\hbar\epsilon_0},
\label{eq:Jph}
\end{equation}
 where, for ease of notation, we have introduced ${\bfg}_{nm}(\omega)\equiv {\bfg}({\bfr}_n,{\bfr}_m;\omega)$ with ${\bf r}_{n/m}$ the positions of $n$-th$/m$-th QD; in addition, ${\bf d}_n = d_n{\bf n}_n$ with ${\bf n}_n$  the unit vector of $n$-th dipole moment, and we have included a pure dephasing term, $L[\sigma_n^{11}]$, on the right hand side of Eq.~(\ref{ms}), with a  dephasing rate $\gamma_n^{\rm '}$, where $L[O]=(O\rho O^{\dagger}-O^{\dagger}O\rho)+{\rm H.c.}$ is the standard
Lindblad superoperator; finally, the last term $L[\sigma_n^+]$ allows for the possibility of an incoherent pump term on each QD$n$ with pump rate $P_n$. In the following calculations, we assume $\gamma'=\gamma_{a}'=\gamma_{b}'$ and $|{\bf d}_n| = 30~$D$\,\approx$ 0.62 e-nm.
As can be seen from the time-dependent integral, the  the LSP bath sampling depends on the system Hamiltonian. 
The time-dependent Pauli matrices are given by $\sigma^{\pm}_n(-\tau) = e^{-{\rm i}H_S\tau/\hbar}\sigma_n^{\pm}e^{{\rm i}H_S\tau/\hbar} = 
\sigma_n^\pm e^{\mp{\rm i}\Delta_{n a}\tau}$, and for resonant QDs, $\Delta_{ab} = 0,$ so we have $\sigma^{\pm}_n(-\tau)=\sigma_n^\pm$. In order to derive Eq.~(\ref{ms}), we have  used the identity $\int d{\bf r}'{\bf G}({\bf r},{\bf r}';\omega)\cdot\varepsilon_{\rm I}({\bf r}',\omega){\bf G}^*({\bf r}',{\bf r}'';\omega)={\rm Im}[{\bf G}({\bf r},{\bf r}'';\omega)]$~\cite{Dung98}.
In a single QNM picture, note that   $J_{\rm ph}(\tau)$ can be divergent; however, in our master equation, 
we calculate $\int_0^tJ_{\rm ph}(\tau)d\tau$, which is convergent.
 In a practical QNM calculation, we compute the frequency integral over a finite bandwidth that covers the QNM resonance, with limits at approximately $\pm$0.5 eV from the QNM resonance frequency. Furthermore, we have  checked that this leads to the correct decay rate from a full dipole numerical simulation.

In Eq.~(\ref{MH}), some non-Markovian effects are captured through the
time integration over the photon reservoir. Applying a 
{\it
second
Markov approximation} (i.e., $t\rightarrow\infty$) for the reservoir bath sampling,  then we obtain the following Markovian master equation
\begin{align}
\frac{\partial\rho}{\partial t} = &{\rm i}[\Delta_{ab}\sigma^{11}_b,\rho]+\sum_n\bigg(
\frac{\gamma^{\rm '}}{2}L[\sigma_n^{11}] + \frac{\gamma_n}{2}L[\sigma_n^-] + \frac{P_n}{2}\nonumber\\
&\times L[\sigma^+_n]-{\rm i}\Delta\omega_n[\sigma_n^{11},\rho]\bigg)+ L_{\rm coup}[\rho],
\label{DynamicsSE}
\end{align}
where we have introduced the QD coupling term 
$
L_{\rm coup}[\rho] ={\rm i}\sum_{n\neq m}\big[(\sigma_n^+\sigma_m^-\rho-\sigma_m^-\rho\sigma_n^+)g_{n m}
-(\rho\sigma_n^+\sigma_m^--\sigma_m^-\rho\sigma_n^+)g^*_{m n}\big]$,
a 
LSP-induced SE rate
 $\gamma_n=2\frac{{\bfd}_n\cdot{\rm Im}[{\bfg}_{nn}(\omega_{n})]\cdot{\bfd}_n}{\hbar\varepsilon_0}$, a photonic Lamb shift $\Delta\omega_{n} = -\frac{{\bfd}_n\cdot{\rm Re}[{\bfg}_{nn}(\omega_{n})]\cdot{\bfd}_n}{\hbar\varepsilon_0}$, and a LSP coupling strength between the QDs, $g_{nm} = \frac{{\bf d}_n\cdot{\bf G}_{n m}(\omega_{m})\cdot{\bf d}_m}{\hbar\varepsilon_0}$. When the QDs are resonant with each other, the coupling term can be simplified to $L_{\rm coup}[\rho]=-{\rm i}\delta_{mn}[\sigma^+_m\sigma^-_n,\rho]+\frac{\gamma_{nm}}{2}(2\sigma_n^-\rho\sigma^+_m
-\sigma^+_m\sigma^-_n\rho-\rho\sigma^+_m\sigma^-_n)$, where $\delta_{mn} = -\frac{{\bfd}_m\cdot{\rm Re}[{\bfg}_{mn}(\omega_{a})]\cdot{\bfd}_n}{\hbar\varepsilon_0}$ describes the coherent coupling between the QDs, and  $\gamma_{nm}=2\frac{{\bf d}_n\cdot{\rm Im}[{\bfg}_{nm}(\omega_{a})]\cdot{\bf d}_m}{\hbar\varepsilon_0}$ describes the incoherent coupling. The master equation (Eq.~(\ref{DynamicsSE})) then simplifies to
\begin{align}
\frac{\partial\rho}{\partial t} = &\sum_{n\neq m}\frac{\gamma_{nm}}{2}(2\sigma_n^-\rho\sigma^+_m
-\sigma^+_m\sigma^-_n\rho-\rho\sigma^+_m\sigma^-_n)\!-\frac{\rm i}{\hbar}[H_{\rm eff},\rho]\nonumber\\
& +\sum_n\bigg(
\frac{\gamma^{\rm '}}{2}L[\sigma_n^{11}] + \frac{\gamma_n}{2}L[\sigma_n^-] + \frac{P_n}{2}L[\sigma^+_n]\bigg),
\label{DynamicsSE1}
\end{align}
where the effective Hamiltonian term is defined as $H_{\rm eff} = \hbar\Delta_{ba}\sigma_b^{11}+\hbar\sum_n \Delta\omega_n\sigma_n^{11}+\hbar\sum_{n\neq m}\delta_{mn}\sigma_m^+\sigma_n^-$.

From  the above master equations, it is clear that the dynamics of the coupled QDs will show a strong positional dependence through the Green function terms, and this is fully captured in the theory.
Unfortunately, the calculation of the Green functions (apart from very simple geometries) is generally  a very difficult and a time consuming process, even with computations carried out on clustered computers. 
In some previous studies, the coupling to the LSP was treated phenomenologically without taking the full detail of the geometry and electromagnetic response into consideration~\cite{twodots1,twodots2}. 
However, for the MNP, recently it has been shown that the Green function can be accurately obtained
in terms of an expansion of the QNM, and for the gold nanorod (and indeed many MNP geometries), a single QNM expansion represents an accurate description of the Green function over broadband frequencies and spatial positions.
For any two spatial points near the MNP, but outside the regime
of Ohmic heating, 
 the dyadic Green function~\cite{comm1} is accurately described through~\cite{quasius1,quasius2}
\begin{align}
{\bfg}_{\rm c}({\bf r}_1,{\bf r}_2;\omega) = \frac{\omega^2}{2\tilde\omega_{\rm c}(\tilde\omega_{\rm c}-\omega)}\tilde{\bf f}_{\rm c}({\bf r}_1)\tilde{\bf f}_{\rm c}({\bf r}_2),
\label{qme}
\end{align}
where $\tilde{\bf f}_{\rm c}({\bfr})$ and $\tilde\omega_{\rm c}$ are the QNM of interest and the correspondent complex eigenfrequency, respectively.
The QNMs are normalized through~\cite{Leung941,Lee99}
\begin{align}
\langle\langle \tilde{\bf f}_{\rm c}|\tilde{\bf f}_{\rm c}\rangle\rangle\!&=\!\lim_{V\rightarrow\infty}\int_V\left(\frac{1}{2\omega}\frac{\partial (\varepsilon({\bf r},\omega)\omega^2)}{\partial \omega}\right)_{\omega=\tilde{\omega}_{\rm c}}\!\!\!\!\tilde{\bf f}_{\rm c}({\bf r})\cdot\tilde{\bf f}_{\rm c}({\bf r})d{\bf r} \nonumber\\
&+ \frac{ic}{2\tilde{\omega}_{\rm c}}\int_{\partial V}\sqrt{\varepsilon({\bf r})}\tilde{\bf f}_{\rm c}({\bf r})\cdot\tilde{\bf f}_{\rm c}({\bf r})d{\bf r},
\label{eq:norm}
\end{align}
where in practise we use a computational volume of about 1.5 micron cubed.
Alternative QNM normalization schemes
are presented in Refs.~[\onlinecite{quasi3,quasiMul}], which have been shown to be  equivalent~\cite{quasius3} to the one above. 
For our MNP resonator, we have assumed, and verified, that there is only one mode in the regime of interest (near the LSP), and for the gold nanorod,  the resonance of the LSP is calculated to be  $\tilde{\omega}_{\rm c}/(2\pi) = \omega_{\rm c}/2\pi + {\rm i}\gamma_{\rm c}/2\pi=
 324.981-{\rm i}16.584~$THz (1.344 - {\rm i}0.0684 eV)~\cite{quasius1} with quality factor $Q=\omega_{\rm c}/2\gamma_{\rm c}\approx9.8$; in order to obtain the QNM numerically, a 6-fs length (Gaussian shape in time domain) spatial plane wave near 325~THz with polarization along the axis of the nanorod is injected, and a run-time Fourier transform with a time window 60~fs is employed; a nonuniform conformal mesh scheme is used, with a mesh size of 1~nm cubed is employed around the nanorod. The  spatial dependence of the mode profile, $|\tilde{\bf f}(x,y,z=0)|$, is shown around the nanorod  in Fig.~\ref{fig:Sf1}(b).

In the calculation of the propagator, we  use the regularized mode $\tilde{\bf F}_{\rm c}({\bf r,\omega})$, since it
allows one to model spatial regimes from the near to far field regimes~\cite{quasius1},
\begin{align}
 {\bf G} \rightarrow {\bf G}^{\rm F}_{\rm c}({\bf r}_1,{\bf r}_2;\omega) =
\frac{\omega^2}{2\tilde\omega_{\rm c}
(\tilde\omega_{\rm c}-\omega)}\tilde{\bf F}_{\rm c}({\bf r}_1,\omega)\tilde{\bf F}_{\rm c}({\bf r}_2,\omega),
  \label{eq:6}
\end{align}
with the {regularized} field  given by $
\tilde{\bf F}_{\rm c}({\bf r},\omega)\equiv \int_{V}{\bf G}^{\rm B}({\bf r},{\bf r'};\omega)\Delta\varepsilon({\bf r'},\omega)\tilde{\bf f}_{\rm c}({\bf r}')d{\bf r}'$, where the volume of the integral is now confined to the nanorod volume and 
$\Delta\varepsilon({\bfr}',\omega)=\varepsilon({\bfr}',\omega)-\varepsilon_B$. The regularized mode has a simple physical interpretation: it is the solution to a scattering problem when the nanorod is excited by the QNM, which ensures the correct
output characteristics in the far field.
 Usually Eq.~(\ref{qme}), which uses the divergent QNM field, gives an excellent approximation to the full Green function as long as the distance between the two positions is no more than a few hundred nm away from the surface of the nanorod; but as the distance becomes sufficiently large, Eq.~(\ref{eq:6}) should be employed to calculate both the propagator and enhancement of LDOS (see Ref.~[\onlinecite{quasius1}] for more details). 

The enhancement of the projected LDOS, in direction ${\bf n}_a$, is defined as
\begin{align}
F_{{\bf n}_a}(\omega)=\frac{{\bf n}_a\cdot{\rm Im}[{\bf G}({\bf r}_a,{\bf r}_a;\omega)]\cdot{\bf n}_a}{{\bf n}_a\cdot{\rm Im}[{\bf G}_{\rm B}({\bf r}_a,{\bf r}_a;\omega)]\cdot{\bf n}_a},
\label{enhancement}
\end{align}
and in terms of the QNM contribution, one simply 
simply replaces ${\bf G}$ by ${\bf G}_{\rm c}$ [i.e., Eq.\,(\ref{qme})].
Figure~\ref{fig:Sf1}(c) shows the comparison between the enhancement of the $x$-projected
(axis of nanorod) LDOS, $F_x$, at 10~nm [${\bf r}_a=(60,0,0)~$nm] away from the nanorod [as in shown in Fig.~\ref{fig:Sf1}(a) by QD$a$], calculated via Eq.~(\ref{qme}) (orange solid line) and  with a full numerical dipole calculation using finite-difference time domain method (FDTD)~\cite{lumerical} (blue dashed line). Clearly the mode expansion technique gives an excellent agreement with the
full-dipole FDTD calculation, and thus includes the LSP reservoir function accurately for use in the presented quantum master equation. The  total SE rate induced by the QNM, including radiative and nonradiative coupling,  is given by 
\begin{equation} 
 \gamma_{a}^{\rm qnm}({\bf r}_a)=\frac{2{{\bf d}_a\cdot{\rm Im}[{\bf G}_{\rm c}({\bf r}_a,{\bf r}_a;\omega_a)]\cdot{\bf d}_a}}{\hbar\varepsilon_0},
\end{equation}
where ${\bf G}_{\rm c}$ is obtained from Eq.~(\ref{qme}).

From the analysis above,  it is clear that two spatially separated QDs could be coupled to each other by the coupling term $L_{\rm coup}[\rho]$ as shown in Eq.~(\ref{DynamicsSE}). In the following, we will also give the  emitted spectrum that can be measured at the detector position ${\bf r}_D$. 
For a system containing $N$ QDs, the spectrum at the position of the detector, ${\bf r}_D$, is given by $S({\bf r}_D,\omega) = \langle ({\bf E}^+_S({\bf r}_D,\omega))^{\dagger}{\bf E}^+_S({\bf r}_D,\omega)\rangle$, with ${\bf E}_S^+({\bf r}_D,\omega) = \frac{1}{\varepsilon_0}\sum_n {\bf G}({\bf r}_D,{\bf r}_{d_n};\omega)\cdot {\bf d}_n\sigma_n^-$ (in the rotating wave approximation).
For continuous wave excitation (e.g., from the incoherent pump field), in the presence of just one QD (e.g., QD $n$), the spectrum is given by  
\begin{align}
S_p({\bf r}_D,\omega) &=
\frac{1}{\varepsilon_0^2}|{\bf d}_n\cdot{\bf G}({\bf r}_n,{\bf r}_D;\omega)|^2 \times \nonumber \\
&\lim_{t\rightarrow\infty}\int_0^{\infty} d\tau e^{-{\rm i}(\omega-\omega_a)\tau}\langle \sigma^+_n(t+\tau)\sigma_n^-(t)\rangle.
\label{spectrum1}
\end{align}
 However for more than one QD, the 
the  power spectrum is derived to be 
\begin{align}
&S_p({\bf r}_D,\omega) = \sum_{n,q} S_0^n(\omega)R_{n}^q({\bf r}_D,\omega)+\nonumber \\
& \ +\sum_{n<m,q}{\rm Re}[S_0^{nm}(\omega) R_{nm}^q({\bf r}_D,\omega)+S_0^{mn}(\omega)R_{mn}^q({\bf r}_D,\omega)],
\label{spectrum}
\end{align}
where $R_{n}^q({\bf r}_D,\omega)=|{\bf d}_n\cdot{\bf G}({\bf r}_n,{\bf r}_D;\omega)\cdot\hat{q}|^2$, and $R_{nm}^q({\bf r}_D,\omega)={\bf d}_n\cdot{\bf G}^*({\bf r}_n,{\bf r}_D;\omega)\cdot\hat{q}\hat{q}\cdot{\bf G}({\bf r}_D,{\bf r}_m;\omega)\cdot {\bf d}_m$ are the generalized propagator factors from the position of the QDs to the  detector, with $m,n= 1, 2, 3, ... $ (or $a,b$ in the present case of two QDs) and $q=x,y,z$; the incoherent spectrum due to the $n$th QD is defined as $S_0^n(\omega)\equiv \frac{1}{\varepsilon_0^2}\lim_{t\rightarrow\infty}{\rm Re}\big[\int_0^\infty d\tau e^{-{\rm i}(\omega-\omega_a)\tau}\langle\sigma_n^+(t+\tau)\sigma_n^-(t)\rangle\big]$; and the cross term due  interference effects between the $n$th and $m$th QD  is given by $S_0^{nm}(\omega) = \lim_{t\rightarrow\infty}\int_0^{\infty} d\tau e^{-{\rm i}(\omega-\omega_a)\tau}\langle \sigma^+_n(t+\tau)\sigma_m^-(t)\rangle$. For convenience, we define the $q$-polarized incoherent spectrum as
\begin{align}
&S_p^q({\bf r}_D,\omega)  = \sum_{n} S_0^n(\omega)R_{n}^q({\bf r}_D,\omega) \nonumber \\ 
 & \ +\sum_{n<m}{\rm Re}[S_0^{nm}(\omega)
 R_{nm}^q({\bf r}_D,\omega)+S_0^{mn}(\omega)R_{mn}^q({\bf r}_D,\omega)],
 \end{align}
 which we will use later to help explain the polarization features of the emitted spectrum.

\section{Quasinormal mode model compared to a Jaynes-Cummings model}
\label{sec:QNMJC}

In the standard JC model both the cavity field and the quantum emitters (e.g., two level atoms) are treated as system operators, which makes the model suitable for studying the physics of strong coupling between the cavity mode and the quantum emitters. To include dissipation into the cavity, the JC model assumes an electromagnetic environment with a Lorentzian spectral density, and this works well for many dielectric cavities. However,  the reservoir function from metals resonators can be highly non-Lorentzian;  moreover, 
it is well known that the plasmonic resonance/spectral density of metallic nanoresonators can change as a function of position around the resonator, which can be probed experimentally by measuring the near field electromagnetic response  at different  positions~\cite{plasmonresonance}. As is shown clearly through Eqs.~(\ref{eq:Jph}),~(\ref{qme}) and~(\ref{eq:6}), the spectral function of the LSP, which is given by the imaginary part of the Green function at the same spatial point, also depends  on the phase of the QNM. Our  general model can actually be used to assess when the JC may work, with a rigorous definition of the coupling parameters, and it can go beyond the Lorentzian lineshape model as well. The JC model, if in a regime of validity, can  explore effects
beyond the 2nd-order Born approximation, e.g., in the strong coupling regime.

\begin{figure}
\centering
\includegraphics[trim=0.70cm 0.1cm .6cm 0.5cm, clip=true,width=.99\columnwidth]{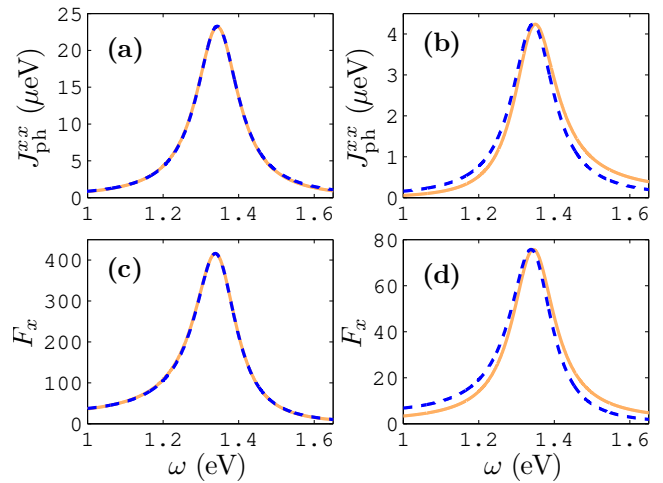}
\caption{(Color online) Spectral function and the enhancement of the LDOS/SE for a single QD position. (a) $J_{\rm ph}^{xx}$  at position ${\bf r}_a$, 10~nm away from the right side of the gold nanorod as shown is Fig.~\ref{fig:Sf1}(b) by the arrow: with the orange (solid) line given by the QNM calculation and the blue (dashed) line is the best Lorentzian fit. (b) QNM calculation of $J_{ph}^{xx}$ at position ${\bf r}_{b'} = (0,0,25)$~nm (orange solid); the blue (dashed) line shows the same Lorentzian fit as in (a), but rescaled in amplitude. (c) Enhancement of the LDOS/SE with the QNM calculation (orange solid) and the Lorentzian function (blue dashed) in (a) for an $x$-polarized QD at ${\bf r}_a$. (d) QNM calculation of $F_x$ for and $x$-polarized dipole at position ${\bf r}_{b'} = (0,0,25)$~nm  (orange solid); and the blue dashed line uses the Lorentzian spectral function from (a).}
\label{fig:Sf12}
\end{figure}

The Green function in Eq.~(\ref{qme}) can be used to obtain  the photon reservoir function [Eq.~(\ref{eq:Jph})], e.g., 
for some dipole position ${\bf r}_a$ (assumed to be near the resonator),
$
{\bfg}_{\rm c}({\bf r}_a,{\bf r}_a;\omega) = \frac{\omega^2}{2\tilde\omega_{\rm c}(\tilde\omega_{\rm c}-\omega)}\tilde{\bf f}_{\rm c}({\bf r}_a)\tilde{\bf f}_{\rm c}({\bf r}_a)$. In a rotating wave approximation, the imaginary part of this function can be written as  \begin{align}
{\rm Im}{\bfg}_{\rm c}({\bf r}_a,{\bf r}_a;\omega)\! =
&\frac{\frac{\omega}{2} \gamma_{\rm c}(\tilde{\bf f}^{\rm R}_{\rm c}({\bf r}_a))^2}
{(\omega_{\rm c}-\omega)^2+\gamma_{\rm c}^2} \left [
1+\frac{{\cal N}({\bf r}_a) (\omega_{\rm c}-\omega)}
{\omega_{\rm c}}
\right ]\!,
\label{qmeRWA}
\end{align}
where $\tilde{\bf f}^{\rm R}_{\rm c}$ and
$\tilde{\bf f}^{\rm I}_{\rm c}$ are the real and imaginary parts of the QNM
function at the dipole position, and 
we have introduced a non-Lorentzian coupling factor defined through
\begin{align}
{\cal N}({\bf r}_a) = \frac{(\tilde{\bf f}^{\rm I}_{\rm c}({\bf r}_a))^2 \omega_{\rm c}}
{\gamma_{\rm c}(\tilde{\bf f}^{\rm R}_{\rm c}({\bf r}_a))^2}.
\end{align}
 
 To better quantify the QNM lineshape, 
 consider a dipole position ${\bf r}_a = (60,0,0)~$nm, 10 nm away from the right side of the metal nanorod, as is shown in Fig.~\ref{fig:Sf1}(b) by the white arrow;
the non-Lorentzian coupling factor is ${\cal N}({\bf r}_a)\approx 1.2$, which mainly leads to a small frequency shift of the resonance frequency (which is easily captured in a Lorentzian function by just moving the resonance frequency).
Thus for this example, the $x$-projected spectral density $J_{\rm ph}^{xx}$ obtained from Eq.~(\ref{eq:Jph}) (orange solid) is well described by a Lorentzian line shape (blue dashed) as shown in Fig.~\ref{fig:Sf12}(a). In general, however,  a position dependent non-Lorentzian spectral density will be obtained around the nanorod; this effect is shown  in Fig.~\ref{fig:Sf12}(b) for $J_{\rm ph}^{xx}$ at the example position ${\bf r}_{b'} = (0,0,25)$~nm, by the solid (orange) line, and a slight blueshift of the resonance peak is also seen which is consistent with the observation in Ref.~[\onlinecite{plasmonresonance}]; the dashed (blue) line is  the  same Lorentzian fit used in the previous case, which clearly shows that the line shape changes as a function of dipole position.  The corresponding enhancement of the LDOS/SE is shown in Fig.~\ref{fig:Sf12}(c). At position ${\bf r}_{b'} = (0,0,25)~$nm, the non-Lorentzian shape of the spectral density influences the enhanced LDOS/SE as is shown in Fig.~\ref{fig:Sf12}(d); the non-Lorentzian coupling factor is now ${\cal N}({\bf r}_a)\approx -2.4$, which has a more dramatic effect on the spectral line shape.  We stress that all the information of the resonance shift and non-Lorentzian spectral function is included in the spatial dependence of the phase factor of the QNM; and this information  naturally comes into the calculations below through the QNM normalization condition  the analytical Green function. Although the Lorentzian fit of the spectral function is valid at certain spatial locations,  the non-Lorentzian spectrum becomes  important when
${\cal N}$ is large enough, and one then requires the
 imaginary part of the QNM as well as the real part~[\onlinecite{quasi3}].

As discussed above, and shown in Fig.~\ref{fig:Sf12}(a), at some positions the spectral density could be well described by a Lorentzian line shape (for certain MNPs), so for QDs at these positions the quantum optical interactions could be approximately described by a dissipative JC model with the following QD-cavity coherent interaction terms (in a rotating wave approximation): ${\bf d}_n\cdot \tilde{\bf f}_{\rm c}({\bf r}_a)a\sigma^+ + {\bf d}_n\cdot \tilde{\bf f}_{\rm c}^*({\bf r}_a)
a^\dagger\sigma^-$;  however, we see that these parameters (and the parameters needed to describe QD-QD interactions) actually  require the QNM technique in order to have a rigorous definition of these coupling parameters and the normalized field.
The single QD-cavity coupling rate will be given by the
usual rate $g$, where $g^2 \propto d_{n}^2|\tilde{\bf f}_{\rm c}({\bf r}_d)|^2$, while
dissipation from the cavity mode  is then usually added through a Lindblad operator that describes only Lorentzian decay. 
  For a dissipative Lorentzian model to work, we find that the QD positions must be  located around high symmetry points within the vicinity of the field antinode points, but far enough away from the metal surface. Even when the JC model approximately works, then the decay rates still have to be obtained as a function of position in general.
This is precisely what the QNM can provide, if the QD position is in a valid Lorentzian decay regime.  

We also caution that the 
JC model still neglects some essential dissipative  coupling processes from the metal environment. 
 For example, the standard JC model does not provide an effective description of the nonradiative and radiative decay processes; such a description of the nonradiative/radiative decay will be necessary  in order to compute important properties such as the quantum yield (or beta factor), e.g., of a single photon source. Below we demonstrate how one can use the QNM technique to achieve the separation of the total decay rate into radiative and nonradiative decay channels.
Moreover, we will also show how one can add in Ohmic losses
in a clear and simple way, which is needed for dipole positions very near  the resonator (e.g., a few nm from the surface)~\cite{quasius2}. Importantly, in our approach, all of these physical rates can be computed
analytically using the QNM theory, as a function of space and frequency.
We describe and exemplify these scattering rates below.

Without the metal nanorod, the background decay rate is simply $\gamma_0 = \frac{2{\bf d}_a\cdot{\rm Im}[{\bf G}_B({\bf r}_a,{\bf r}_a;\omega_a)]\cdot{\bf d}_a}{\hbar\varepsilon} = \frac{d_a^2\omega_a^3n_B}{3\hbar\varepsilon_0\pi c^3}$.
While the nonradiative decay rate from the QNM is obtained from~\cite{nr}
 \begin{equation}
 \gamma_{a}^{\rm nrq}({\bf r}_a)=\frac{2}{\hbar\omega_a\varepsilon_0}\int_{V_{\rm MNP}}\!{\rm Re}[{\bf j}({\bf r})\cdot{\bf G}_{\rm c}^*({\bf r},{\bf r}_a;\omega_a)\cdot{\bf d}_a]d{\bf r},
\end{equation}
where  ${\bf j}({\bf r})=\omega_c\varepsilon_I({\bf r},\omega_a){\bf G}_{\rm c}({\bf r},{\bf r}_a;\omega_a)\cdot{\bf d}_a$ is the induced current density in the nanorod (MNP) at position ${\bf r}_a$.
Thus the radiative decay rate from the QNM is given by
\begin{equation}
\gamma_{a}^{\rm rq}({\bf r}_a)=\gamma_{a}^{\rm qnm}({\bf r}_a)
- \gamma_{a}^{\rm nrq}({\bf r}_a).
\end{equation}
 In addition, the quasistatic  decay rate can be obtained from
\begin{equation}
\gamma_{a}^{\rm stat}({\bf r}_a)=\frac{2{{\bf d}_a\cdot{\rm Im}[{\bf G}^{\rm qs}({\bf r}_a,{\bf r}_a;\omega_a)]\cdot{\bf d}_a}}{\hbar\varepsilon_0},
\end{equation}
 with ${\bf G}^{\rm qs}({\bf r}_a,{\bf r}_a;\omega_a)=\mp{\bf G}_{\rm B}({\bf r}_a',-{\bf r}_a';\omega_a)\frac{\varepsilon(\omega_a)-\varepsilon_B}
{2(\varepsilon(\omega_a)+\varepsilon_{B})}$~\cite{quasius1,quasistat} ($\mp$ is for $s$/$p$-polarized dipoles, respectively). Consequently, the total nonradiative decay is given by $\gamma_{\rm nr} = \gamma_a^{\rm nrq}+\gamma_a^{\rm stat}$. As is shown above, all of the decay rates are highly position dependent, but once the QNM is calculated, the decay rates at different positions can be computed immediately.
 
  \begin{figure}
\centering\includegraphics[trim=2.0cm 1.2cm 3.0cm 4.0cm, clip=true,width=.999\columnwidth]{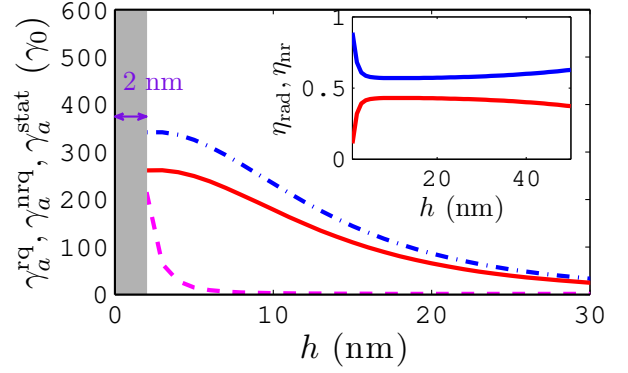}
\caption{(Color online)  Decay rates (in units of the homogeneous space  radiative decay rate $\gamma_0$) of an $x$-polarized QD$a$ induced by quasi-static interaction $\gamma^{\rm stat}_a$ (magenta dashed), radiative contribution of the QNM $\gamma^{\rm rq}_a$ (red solid), and the non-radiative contribution of the QNM $\gamma_a^{\rm nrq}$ (blue chain), as a function of $h$ [${\bf r}_{a}=(50~{\rm nm}+ h,0,0)$] away from the nanorod;
inset shows the radiative coupling factor, $\eta_{\rm rad}$ (lower red solid), and non-radiative coupling factor, $\eta_{\rm nr}$ (upper blue solid) of the decay rate.}
\label{fig:srate2}
\end{figure} 
Figure~\ref{fig:srate2} shows the decay rates as a function of distance, $h$, away from the surface of the metal nanorod along the $x$-axis; where we see that, at extremely small distances $h$, the system is in quasi-static regime where the Ohmic heating effect due to $\gamma_a^{\rm stat}$ (magenta dashed) is strong; as $h$ becomes larger, all of the decay rates decrease, but the quasi-static decay rate decreases much faster than the others; the inset of Fig.~\ref{fig:srate2} shows the radiative coupling factor, $\eta_{\rm rad} = \frac{\gamma_a^{\rm rq}}{\gamma_a^{\rm qnm}+\gamma_a^{\rm stat}}$ (lower red solid), and nonradiative coupling factor, $\eta_{\rm nr} = \frac{\gamma_a^{\rm stat}+\gamma_a^{\rm nrq}}{\gamma_a^{\rm qnm}+\gamma_a^{\rm stat}}$ (upper blue solid), as function of $h$ in the near field regime. We see that the
radiative output coupling efficiency is below 50\%, though this can be increased
to about 60\% or greater using a dimer configuration~\cite{quasius2}.
While it is not clear how to include such processes in a JC model, which would be further complicated by having different parameters at different QD positions, they could certainly help improve and guide such simpler models. More details on such an approach will be reported in a future publication when we will also explore effects beyond a second-order Born approximation.

With regards to computing the spectrum in a JC model, 
the spatially integrated far-field spectrum from the cavity operator is usually given by (assuming a rotating frame as the cavity frequency) $S_{\rm cav}(\omega)\propto \kappa \lim_{t\rightarrow\infty}\int_0^{\infty} d\tau e^{-{\rm i}(\omega-\omega_c)\tau}\langle a^\dagger(t+\tau)a(t)\rangle$; however this assumes that the output coupling rate via the cavity (in this case the LSP) is purely radiative. For a metal resonator, one must include quenching effects by solving the input/output scattering problem, which is exactly what the Green function solution has  done. In this way one can compliment the JC model by computing the spatially dependent output spectrum from the QD system operator dynamics and the medium electric field operators, so that  
$S_{\rm cav}({\bf r},\omega)\propto |{\bf d}\cdot {\bf G}_{\rm c}({\bf r},{\bf r}_d;\omega)|^2 \lim_{t\rightarrow\infty}\int_0^{\infty} d\tau e^{-{\rm i}(\omega-\omega_c)\tau}\langle \sigma^+(t+\tau)\sigma^-(t)\rangle$. Furthermore, one could obtain the spatially averaged output spectrum (e.g., in the far-field) from $S_{\rm cav}^{\rm rad}(\omega)\propto \eta_c \kappa \lim_{t\rightarrow\infty}\int_0^{\infty} d\tau e^{-{\rm i}(\omega-\omega_c)\tau}\langle a^\dagger(t+\tau)a(t)\rangle$, where $\kappa=\gamma_{\rm c}$ and the radiative output coupling factor associated with the QNM is obtained from $\eta_c = \frac{\gamma_a^{\rm rq}}{\gamma_a^{\rm qnm}}$.

To summarize this section, we have discussed how our model can go well beyond the standard JC model while facilitating the simpler JC models in certain spatial regimes. To the extent that the approximate JC could be valid, one still has to obtain the coupling parameters from a model such as ours, and then carefully include quenching effects into any calculation of emitted fields far away from the system resonator. Thus our model can be used to guide and help the simpler JC models in certain regimes as well.

\section{Results and applications}
\subsection{Localized plasmon  induced SE from a single excited QD near the nanorod}
\label{sec:resulta} 
Metal nanoparticles enhance the SE rate of excited single QDs
due to the coupling with  the LSP (QNM). The LDOS at positions around the nanorod changes rapidly in space  compared to the homogeneous dielectric structure, 
and thus the SE rate of a QD around the nanorod can be significantly 
changed~\cite{quasi3,PhilipACS,quasius1,quasius2}, as is shown in Fig.~\ref{fig:Sf1}(c).
 In this section, we  present an analysis of the SE dynamics of a single QD on resonance with the LSP ($\omega_{a} = \omega_{\rm c})$; without loss of generality, we take the case of a QD polarized along the $x$-axis at position ${\bf r}_a$, 10~nm away from the nanorod as shown schematically in Fig.~\ref{fig:Sf1}(a) (QD$a$). 
 For the single-QD nanorod system, without an incoherent pump field (i.e., $P_a=0$), the  non-Markovian master (Eq.~(\ref{ms})) becomes
\begin{align}
\frac{\partial  \rho}{\partial t}=&\int_0^td\tau\left(\,[\sigma^-\rho\sigma^+-\sigma^+\sigma^-\rho ]J_{\rm ph}(\tau)+{\rm H.c.}\right)\nonumber\\
& + \frac{\gamma^{\rm '}}{2}L[\sigma^{11}],
\label{SE}
\end{align}
where we have explicitly used the result $\sigma^{\pm}(-\tau)=\sigma^{\pm}$ since $H_S=0$ and the kernel function is given by $J_{\rm ph}(\tau)=J_{\rm ph}^{aa}(\tau)$. Note the population decay is not affected by pure dephasing here so we can  neglect pure dephasing for this single QD radiative decay study.

\begin{figure}
\centering\includegraphics[trim=0.6cm 1.1cm 5.0cm 4.2cm, clip=true,width=.95\columnwidth]{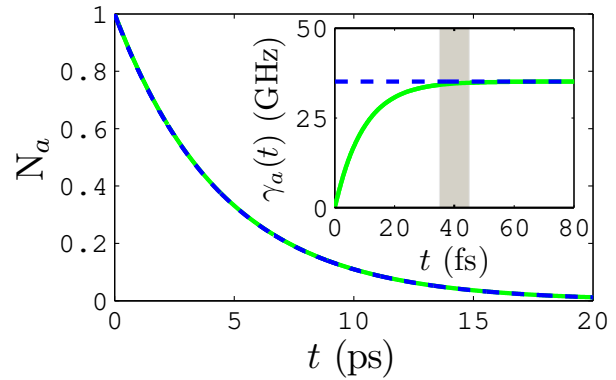}
\caption{(Color online) Population dynamics of an excited single QD, $N_a$, for an $x$-polarized dipole with $|{\bf d}|=~$30 D located at, ${\bf r}_a=(60,0,0)~$nm, 10~nm away from the nanorod as shown in Fig.~\ref{fig:Sf1}(a) by QD$a$. Exponential decay with $\gamma_a = \frac{2{\bf d}_a\cdot{\rm Im}[{\bf G}({\bf r}_a,{\bf r}_a;\omega_a=\omega_c)]\cdot{\bf d}_a}{\hbar\varepsilon_0}$ shown by the blue dashed line, and the green solid line is the full non-Markovian dynamics given by Eq.~(\ref{SE}). The inset shows the effective exponential decay rate, $\gamma_a(t)$, calculated with the full decay dynamics (green solid), and $\gamma_a^0$ (blue dashed); at the crossover region, around 40~fs ($\approx 2\pi/\gamma_c$), $\gamma_a(t)$ agrees with $\gamma_a^0$ within 2\% as shown by the light grey area). }
\label{fig:SED}
\end{figure}

We assume here that the QD is initially excited.
The QD population decay,  ${\rm N}_a = \rho_{ee} = \langle e|\rho|e\rangle$,  using Eq.~(\ref{ms}), is shown in Fig.~\ref{fig:SED} by the light (green) solid line; the  dashed (magenta) line shows the result of a Markovian exponential decay with the  rate, $\gamma_a = \frac{2{\bf d}_a\cdot{\rm Im}[{\bf G}({\bf r}_a,{\bf r}_a;\omega_a)]\cdot{\bf d}_a}{\hbar\varepsilon_0}$ given by Fermi's golden rule with the on-resonant projected LDOS. 
The inset to Fig.~\ref{fig:SED} shows that the SE dynamics  is recovered by  Fermi's golden rule after a characteristic timescale of about 40~fs (shown in the light gray region); this time scale agrees very well with the corresponding lifetime of the LSP, $\tau_{c}\approx2\pi/\gamma_{\rm c}$.

\subsection{Localized  plasmon  induced coupling between two spatially separated QDs in homogeneous background}
\label{sec:resultsb}
\begin{figure}
\centering\includegraphics[trim=0.6cm 1.8cm 1.0cm 2.0cm, clip=true,width=.999\columnwidth]{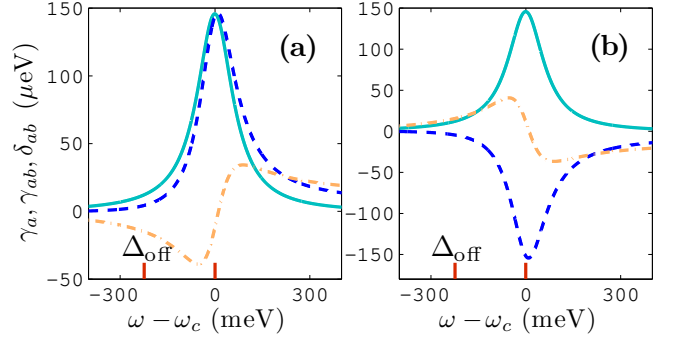}
\caption{(Color online) The various scattering rates as the function of detuning from the resonance  of the LSP ($\omega_c$) for two resonant QDs with  QD$a$ at ${\bf r}_a=(60,0,0)~$nm (i.e., 10~nm away from the nanorod surface). (a) $\gamma_{aa}$ (cyan solid), $\gamma_{ab}$ (blue dashed), and $\delta_{ab}$ (orange dash-dot) with QD$b$ at ${\bf r}_{b} = (-60,0,0)~$nm; the left and right short (red) vertical lines are for later reference when we choose QD detunings of  $\Delta=\omega_a-\omega_{\rm c}=-224~$meV $\equiv\Delta_{\rm off}$, and $\Delta=0~$meV, respectively. (b) same as (a) but with ${\bf r}_{b'} = (-45,0,23)~$nm [see Fig.~\ref{fig:Sf1}(a)]. }
\label{fig:srate1}
\end{figure}

For two spatially separated QDs  located around the nanorod (as is shown in Fig.~\ref{fig:Sf1}(a) by QD$a$ and QD$b$), these can be effectively coupled to each other by exchanging photons via the LSP; in the absence of a pump field (i.e., $P_{n}=0$), the non-Markovian master equation becomes
\begin{align}
\frac{\partial \rho}{\partial t}=&\int_0^t\bigg(\,\sum_{n,m}[\sigma_n^-e^{{\rm i}\Delta_{n a}\tau}\rho\sigma_m^+-\sigma^+_n\sigma_m^-e^{{\rm i}\Delta_{m a}\tau}\rho ]J_{\rm ph}^{n l}(\tau)\nonumber\\
&+{\rm H.c.}\bigg)d\tau + \sum_n\frac{\gamma^{\rm '}_n}{2}L[\sigma^{11}_n]+{\rm i}[\Delta_{ab}\sigma_b^{11},\rho].
\label{COUPLEDME}
\end{align}
 Unless stated otherwise, we will assume the two QDs are resonant with each other ($\Delta_{ab}=0$), but may be off resonant with the LSP, where $\omega_a=\omega_b= \omega_{\rm c}+\Delta$; however, later we also study the case with different QD resonance frequencies [e.g., in Fig.~\ref{fig:fPL3}(d)]. Here the intercoupling between the QDs depends on the  projected cross density of optical states (CDOS), $\varrho_{ab}\equiv \varrho({\bf r}_a,{\bf r}_b;\omega) \equiv {\bf n}_a \cdot {\rm Im}[{\bf G}({\bf r}_a,{\bf r}_b;\omega)] \cdot {\bf n}_b$ via $J^{ab}_{\rm ph}$, which gives one part of the characteristic coupling strength between the QDs mediated by the electromagnetic environment of the nanorod. Since the Green function in use is the retarded Green function,
the real and imaginary parts are related to each other through the Kramers-Kronig relation. The real part of the Green function between the two QDs yields the coherent coupling, $\delta_{ab}(\sigma_a^+\sigma^-_b+\sigma_b^+\sigma_a^-)$, while the imaginary part  gives the incoherent coupling, $\sum_{n\neq m}\frac{\gamma_{ab}}{2}(2\sigma_{n}^-\rho\sigma_m^+-\sigma_m^+\sigma_n^-\rho-\rho\sigma_m^+\sigma_n^-)$; the relevant coupling strengths are shown in Fig.~\ref{fig:srate1} with the coherent coupling ($\delta_{ab}$) and the incoherent coupling strength ($\gamma_{ab}$) given by the chain (orange) and dashed (blue) lines, respectively, and the solid (cyan)  line shows $\gamma_a$. It can be seen from Fig.~\ref{fig:srate1}(a), for $x$-polarized QDs at positions ${\bf r}_{a/b}=(\pm60,0,0)~$nm, that $\delta_{ab}$ may dominate over $\gamma_{ab}$ and $\gamma_a$ when $\Delta=\Delta_{\rm off}$, where 
$\Delta_{\rm off}$ is some offset frequency from the real part of the LSP resonance; however, when  QD$b$ is $z(y)$-polarized, there is almost no coupling between the QDs;
but due to the complex position-dependent 
 polarization characteristics of the LSP, the off-diagonal element of both the projected CDOS, $\varrho_{ab}(\omega)$, and the real part of the Green function are non-zero; consequently, QDs with different polarization can be effectively coupled to each other for certain QD positions. Figure~\ref{fig:srate1}(b) shows that an $x$-polarized QD$a$ at ${\bf r}_a= (60,0,0)~$nm could be effectively coupled to a $z$-polarized QD$b$ at ${\bf r}_{b'}=(-45,0,23)~$nm.
Below, we will also look at the effect of the coherent exchange interactions in the presence of QD pure dephasing.

\begin{figure}
\centering\includegraphics[trim=0.6cm 0.6cm 0.6cm 4.5cm, clip=true,width=.999\columnwidth]{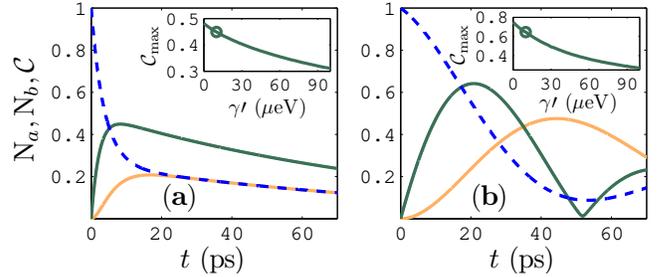}
\caption{(Color online) Dynamics of two resonant QDs  with $x$-polarized excited QD$a$ and unexcited QD$b$, $|eg\rangle$, at ${\bf r}_a = (60,0,0)~$nm, ${\bf r}_b = (-60,0,0)~$nm, respectively. (a) Time evolution of the exciton population of QD$a$/QD$b$, $N_{a/b}$ (blue dashed/orange solid), and entanglement $\cal{C}$ (dark green solid) with pure dephasing rate $\gamma'=10~\mu$eV for $\Delta=0$. The inset shows $\cal{C}_{\rm max}$ as a function of $\gamma'$ and the circle shows the position at which the dynamics  is studied. (b) same as (a) except for $\Delta=\Delta_{\rm off}$.}
\label{fig:coupled}
\end{figure} 

We first assume that the two QDs are initially in a separable state $|eg\rangle$ (with the first argument for QD$a$ and the second one for QD$b$); as a result of the coherent coupling, non-classical correlations will be induced between the QDs, and the quantum correlations approach a maximum  value at
some characteristic interaction time, which  eventually decays to zero  due to the decoherence caused by the strong dissipation and the pure dephasing of the system. As a measure of the nonlocal quantum correlations between the separated QDs, we use the concurrence
${\cal C}$, which is obtained from the eigenvalues of the flipped density matrix~\cite{concurrence}, and its maximum value  in the evolution is denoted as ${\cal C}_{\rm max}$. The exciton population of  QD$n$ is defined as ${\rm N}_{n}=\langle e_n|{\rm tr}_m\rho|e_n\rangle$ with $n\neq m$. Figure~\ref{fig:coupled}(a) shows the dynamics of $\cal{C}$ for $x$-polarized QDs on resonance with the LSP ($\Delta=0$) at positions ${\bf r}_{a/b}=(\pm60,0,0)~$nm [shown in Fig.~\ref{fig:Sf1}(a) by the dark brown ellipsoids] by the green (dark) solid line for $\gamma'=10~\mu$eV; the exciton populations $N_{a/b}$ are shown by the blue dashed line and orange (light) solid line, respectively; the maximum ${\cal C}_{\rm max}$  as a function of the pure dephasing rate $\gamma'$ is shown in the inset and, for this case, ${\cal C}_{\rm max}$ is always less than 0.5 in agreement with previous work for entangled atoms in weakly coupled reservoirs~\cite{entanglement}. As shown in Fig.~\ref{fig:srate1}(a), when the QDs are on resonance with the LSP, the incoherent coupling rates are much larger than the coherent coupling rate $|\gamma_{ab}|,|\gamma_{a}|\gg|\delta_{ab}|$, but as they are detuned away from the LSP resonance, the coherent coupling strength begins to dominate over the incoherent coupling. Figure~\ref{fig:coupled}(b) shows the same calculation as \ref{fig:coupled}(a) but with $\Delta=\Delta_{\rm off}$, and now we see that ${\cal C}_{\rm max}$ could be much larger than the previously  limit of 0.5~\cite{entanglement}; indeed our calculations show that it could be even larger than 0.7 if the pure dephasing rate were smaller; in addition, we see that the concurrence exhibits an oscillating behaviour, which is similar to that of the coherent system indicating that the two-QD are  effectively coupled through the LSP-induced photon exchange.

For the detuning value of $\Delta=\Delta_{\rm off}$, the incoherent rates are around $\gamma_{a/b}\approx14~\mu$eV, the incoherent coupling rate $\gamma_{ab}\approx 6~\mu$eV, while the coherent coupling strength is  $\delta_{ab}\approx 17.5~\mu$eV;  in contrast, for  the on resonance case (i.e., $\Delta=0$), we have $\gamma_{a/b}\approx148~\mu$eV,$\gamma_{ab} \approx 146~\mu$eV, and $\delta_{ab}\approx 14~\mu$eV. It can be seen that the relative coherent coupling strength with a finite detuning ($\Delta=\Delta_{\rm off}$) is much larger than it is at $\omega_{\rm c}$  (neglecting $\gamma'$). However, as $\gamma'$ increases, the effective coherent coupling strength for $\Delta=\Delta_{\rm off}~$ decreases much faster than for $\Delta=0$. Thus when $\gamma'=0$, $\cal{C}_{\rm max}$ for $\Delta=\Delta_{\rm off}~$ is larger than that for $\Delta=0$, but it decreases faster as well since  $\gamma'$ increases---as shown in the insets of Fig.~\ref{fig:coupled}.

\begin{figure}
\centering\includegraphics[trim=0.5cm 1.3cm 1.7cm 0.0cm, clip=true,width=.999\columnwidth]{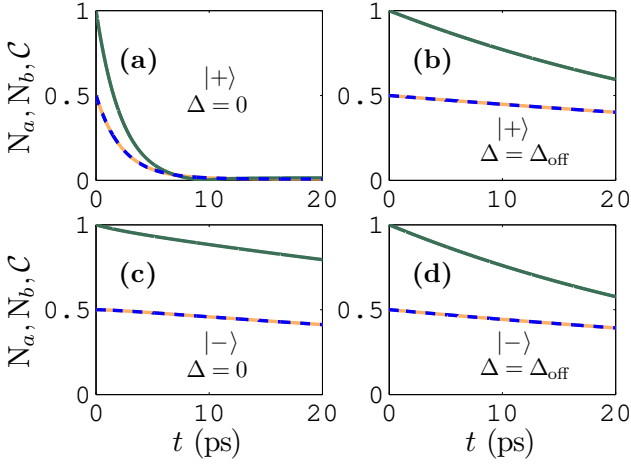}
\caption{(Color online) Population decay dynamics  of the initial Bell initial states $|\pm\rangle$ for both $x$-polarized QDs with the same resonance frequency,  at positions ${\bf r}_{a/b}=(\pm60,0,0)~$nm, respectively; the pure dephasing rate is $\gamma'=10~\mu$eV. (a) For the initial state $|+\rangle$ with $\Delta=0$ the dark green solid line is $\cal{C}$, and the blue dashed/orange solid lines show the exciton population of QD$a$/QD$b$, respectively; (b) same as (a) but with $\Delta=\Delta_{\rm off}$ (see Fig.~\ref{fig:srate1}); (c) same as (a) but with the initial state $|-\rangle$;  (d) same as (c) but with $\Delta=\Delta_{\rm off}$.
}
\label{fig:Bell}
\end{figure} 

It is also demonstrated in Fig.~\ref{fig:coupled} that, due to the presence of QD$b$, the decay of QD$a$ in the long time limit slows down. This effect can be explained through the effective Hamiltonian, $H_{\rm eff}$, which in the absence of dissipation results in four eigenstates $|ee\rangle,|gg\rangle,|\pm\rangle=\frac{1}{\sqrt{2}}(|eg\rangle\pm|ge\rangle)$. The initial state, $|eg\rangle$, lies in the sub-space composed of $|\pm\rangle$
which gives the superradiant and subradiant emission depending on the relationship among the enhanced SE rate, $\gamma_{a}$, and the incoherent coupling, $\gamma_{ab}$. 
Thus the decay of the excited QD may be enhanced at the beginning ($t\rightarrow 0$) due to the faster decay of {the component of superradiant state in the initial state}; 
 while at long times, the dynamics is dominated by the slower decay of the component of subradiant state in the initial state, which gives a suppressed emission if there is a considerable amount of the subradiant component in the initial state. 

Figure~\ref{fig:Bell} shows the dynamics of the resonant QDs located symmetrically at ${\bf r}_{a/b}=(\pm60,0,0)~$nm with the initial states $|\pm\rangle$, with a pure dephasing rate $\gamma'=10~\mu$eV. It is shown in Fig.~\ref{fig:Bell}(a), that when the QDs are on resonant with the LSP, $|+\rangle$ is the superradiant state ($\gamma_{a/b}\approx\gamma_{ab}\gg\gamma'$), and $N_{a/b}$ (blue dashed/orange solid) decay twice as fast than 
QD$a$ alone.  The dynamics with the initial state $|-\rangle$ is shown is Fig.~\ref{fig:Bell}(c), which is now the  subradiant state. With a detuning of $\Delta_{\rm off}$, we have $\gamma_{a/b}\approx2\gamma_{ab}$,  which are much less than $\gamma_a$ at $\Delta=0$, so there is not much difference between the superradiant and subradiant states as is shown in Figs.~\ref{fig:Bell}(b)-(d).
To establish if there are any non rotating-wave effects not captured by our master equation approach, 
 we have also checked that an exact wavefunction method based on the schr\"{o}dinger equation~\cite{nonr1,nonr2} (with no rotating wave approximation, but restricted to weak excitation with no pure dephasing) gives the same solution as above with no noticeable difference.

We stress that with our QNM formulation, one does not need to calculate additional Green function simulations for different QD positions, which makes the approach   convenient for exploring the position-dependent behaviour of QDs (as we have demonstrated earlier for the position dependent decay rates). For the initial state $|eg\rangle$, numerical calculations (with $\Delta_{ab}=\Delta=0$, $\gamma'=10~\mu$eV) show that the maximum achievable entanglement, $\cal{C}_{\rm max}$, is not a monotonic function of distance $h$ from the QDs to the both sides of the nanorod, ${\bf r}_{a/b} = (\pm 50\pm h,0,0)~$nm. It is found that, at first $\cal{C}_{\rm max}$ increases as $h$ becomes larger, and reaches its maximum around $h=10$~nm; then, it decreases as $h$ increases further; for example,  ${\cal{C}}_{\rm max}(h=2~{\rm nm})\approx 0.38$, ${\cal{C}}_{\rm max}(h=10~{\rm nm}) \approx 0.45$, and ${\cal{C}}_{\rm max}(h=18~{\rm nm}) \approx 0.39$. This could be explained by analyzing the radiative and nonradiative decay rates earlier. 
As is shown in Fig.~\ref{fig:srate2}, at extremely small $h$, the system is in  the quasi-static coupling regime where  Ohmic losses  due to $\gamma_a^{\rm stat}$ (magenta dashed) are strong; as $h$ becomes larger, the Ohmic losses becomes smaller and the effective coupling between the QDs becomes larger and thus $\cal{C}_{\rm max}$ increases; however,  as the spatial distance increases further, the effective coupling strength becomes weaker and weaker with respect to the pure dephasing rate, $\gamma'$ , which causes $\cal{C}_{\rm max}$ to decrease again.

\subsection{Emitted spectrum from an incoherent pump}
\label{sec:resultsc}
As is analysed in Sec.\ref{sec:theory} and shown explicitly in Sec.~\ref{sec:resultsb}, two spatially separated QDs can be effectively coupled to each other due to the characteristics of the CDOS (incoherent coupling) and the real part of the Green function (coherent coupling) of the nanorod. In the following, we will concentrate on the  spectrum that can be measured using excitation from an incoherent pump field.

\begin{figure}
\centering\includegraphics[trim=0.2cm 0.8cm 0.4cm 0cm, clip=true,width=.999\columnwidth]{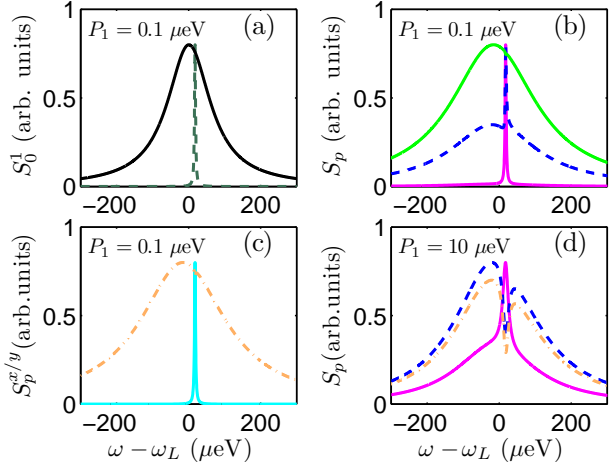}
\caption{(Color online) Incoherent spectra for QDs at ${\bf r}_{a/b}=(\pm60,0,0)~$nm, respectively ($\omega_{a/b}=\omega_c$) with $\gamma'=1~\mu$eV. (a) $S_0^1$ with $P_1=0.1~\mu$eV: the black solid/green dashed are results with/without the presence of QDs, respectively; (b) Incoherent spectra $S_p$ at position ${\bf r}_D=(0,0,0.1/0.3/2)~\mu$m (magenta/blue dashed/green) with $P_a = 0.1~\mu$eV. (c) Polarization dependent spectra $S_p^x$ (orange chain) and $S_p^z$ (cyan) at ${\bf r}_D=(0,0,0.1)~\mu$m with $P_a = 0.1~\mu$eV. (d) $S_P$ at position
${\bf r}_D=(0,0,0.1/0.3)~\mu$m (magenta/blue dashed); $S_p^x$ (orange chain) at ${\bf r}_D=(0,0,0.1)~\mu$m with $P_a = 10~\mu$eV.}
\label{fig:fPL1}
\end{figure} 

\begin{figure}
\centering\includegraphics[trim=0.2cm 2.0cm 0.4cm 2.8cm, clip=true,width=.999\columnwidth]{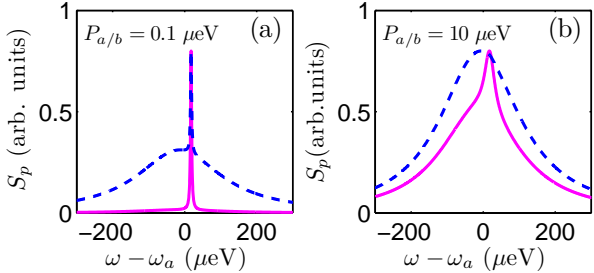}
\caption{(Color online) Incoherent spectra, $S_P$ at position ${\bf r}_D=(0,0,0.1/0.3)~\mu$m (magenta/blue dashed) for QDs at ${\bf r}_{a/b}$ respectively ($\omega_{a/b}=\omega_c$) with $\gamma'=1~\mu$eV. (a) $P_{a/b} = 0.1~\mu$eV. (d) $P_{a/b} = 10~\mu$eV.}
\label{fig:fPL0}
\end{figure}

For our first incoherent pump investigation, we assume both the two $x$-polarized QDs are resonant with the LSP of the nanorod ($\Delta_{ab}=\Delta=0$), and they are symmetrically located at 10~nm (${\bf r}_{a/b}$) away from the both sides of the nanorod ($\gamma_a=\gamma_b$, $\Delta\omega_a=\Delta\omega_b$) as is shown in Fig.~\ref{fig:Sf1}(a). We first assume only QD$a$ is incoherently pumped ($P_b=0$), and we will compare this result with the spectrum emitted
when both QDs are incoherently excited. The emitted spectra  are shown in Fig.~\ref{fig:fPL1},  with $\gamma'=1~\mu$eV and $P_b=0$. Without the presence of QD$b$, the  bare spectrum of QD$a$, $S_0^a$, is shown in Fig.~\ref{fig:fPL1}(a) by the black solid line, and its full-width at half maximum (FWHM) is much larger than for a homogeneous medium due to the LSP coupling; the green (dark) dashed line is $S_0^a$ from QD$a$ including the presence of QD$b$. The linear spectrum, $S_p$, at difference positions are shown in Fig.~\ref{fig:fPL1}(b) for ${\bf r}_D=(0,0,0.1/0.3/2)~\mu$m (magenta solid,  blue dashed, green solid). It is interesting that the spectrum in the near field regime, ${\bf r}_D=(0,0,0.1)~\mu$m, shows a sharp spectral peak; but as the detector position is located further away from the nanorod, at ${\bf r}_D=(0,0,0.3)~\mu$m, a broadened peak with a sharp peak located at the same spectral position as in the near field is observed; as the spectrum propagates  to the far field regime, at ${\bf r}_D=(0,0,2)~\mu$m, then the sharp peak develops into a dip, which indicates that there is interference between the sharp peak and the broad resonance which is a  Fano resonance effect. From the analysis in Sec.~\ref{sec:resultsb}, $|\pm\rangle$ are eigenstates of the $H_{\rm eff}$, which are the  superradiant and subradiant states, respectively. The sharp peak is the result of decay from $|-\rangle$ to $|gg\rangle$, and the broad peak is the decay from $|+\rangle$ to $|gg\rangle$, while $S_p$ is the total contributions from the two including interference effects; the separation between the peaks is given by $2\hbar |\delta_{ab}|$ in the linear regime, and the asymmetry of the position with respect to $\omega_a$ depends on the  induced Lamb-shift $\Delta\omega_{a/b}$. In Figs.~\ref{fig:fPL1}(a) and (b), the bare spectrum is almost the same as the sharp peak, which means the population of $|-\rangle$ is much larger than $|+\rangle$ in the linear regime where $P_a\ll\gamma_{a/b},\gamma',\gamma_{ab}$. In fact, under this situation, the rate equations of $\rho_{++}$, and $\rho_{--}$ are simply given by
\begin{subequations}
\begin{align}
\frac{d\rho_{++}}{d t} = &\frac{\gamma'}{2}(\rho_{--}-\rho_{++})+\gamma(\rho_{ee}-\rho_{++})
\nonumber \\
&+\frac{P_a}{2}(\rho_{gg}-\rho_{++})
+\gamma_{ab}(\rho_{ee}-\rho_{++}),\\
\frac{d\rho_{--}}{d t}=&\frac{\gamma'}{2}(\rho_{++}-\rho_{--})+\gamma(\rho_{ee}-\rho_{--})\nonumber \\
&+\frac{P_a}{2}(\rho_{gg}+\rho_{--})
+\gamma_{ab}(\rho_{--}-\rho_{ee}).
\end{align}
\end{subequations}
Since at the steady state $\rho_{gg}\approx1$, $\rho_{ee}\approx0$ {for a small pump rate, $P_a$}, then the ratio of steady state population $\rho_{--}$ to $\rho_{++}$ is given by $\frac{\rho_{--}}{\rho_{++}}\approx\frac{\gamma'+\gamma+\gamma_{12}}{\gamma'+\gamma-\gamma_{12}}\gg 1$.

Figure~\ref{fig:fPL1}(c) shows the polarization dependent spectra $S_p^z$ (cyan solid), and $S_p^x$ (orange chain); it is seen that the sharp peak is mainly $z$-polarized, while the broad peak is primarily $x$-polarized. So, at the far field, the $S_p$ displays mainly the broad peak as a result of the dipole radiation that is observed in Fig.~\ref{fig:fPL1}(b). In the presence of a nonlinear pump field, with $P_a=10~\mu$eV, the computed spectrum $S_p$, is shown in Fig.~\ref{fig:fPL1}(d) at ${\bf r}_D=(0,0,0.1)~\mu$m by the magenta solid line, and the orange chain is the $x$-polarized spectrum $S_p^x$.

We next consider both QDs incoherently excited by the same pump field, with $P_{a}=P_{b}$.
In the linear regime (weak pump limit), 
 we get basically the same result as the case with one QD incoherently pumped, as shown in Fig.~\ref{fig:fPL0}(a). As long as the pump rate is small, the incoherent spectrum is similar with one or two pump fields.
However, as the pump field is increased, the power broadening with two QDs excited
is notably larger, as depicted
in Fig.~\ref{fig:fPL0}(b). While there are many pumping scenarios that we could study, in what follows, we will concentrate on the case that only QD$a$ is incoherently excited.

\begin{figure}
\centering\includegraphics[trim=0.3cm 2.1cm 0.4cm 3.0cm, clip=true,width=.999\columnwidth]{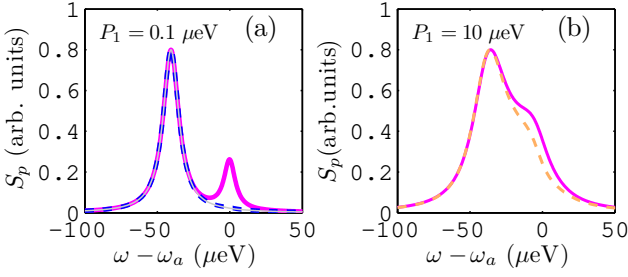}
\caption{(Color online) Incoherent spectra for resonant QDs ($\omega_{a/b}=\omega_c+\Delta_{\rm off}$) at positions ${\bf r}_{a/b} = (\pm60,0,0)~$nm respectively with $\gamma'=1~\mu$eV. (a) $S_p$ at position ${\bf r}_D = (0,0,0.1/0.3/2)~\mu$m (thick magenta solid/thick blue dashed/thin grey solid) with $P_a=0.1~\mu$eV. (b) $S_p$ (magenta) and $S_p^x$ (orange dashed) at position ${\bf r}_D = (0,0,0.1)~\mu$m with $P_a=10~\mu$eV.}
\label{fig:fPL2}
\end{figure}

As we detune the QDs from the LSP resonance, using $\Delta = \Delta_{\rm off}$, a double-peak is observed for the total spectrum, $S_p$,  at ${\bf r}_D=(0,0,0.1)~\mu$m as is shown in Fig.~\ref{fig:fPL2}(a) by the magenta solid line; as before,  the right (higher frequency) peak is suppressed as we evolve to the far field regime. The corresponding high pump spectrum $S_p$ is shown in Fig.~\ref{fig:fPL2}(b) by the magenta solid line at ${\bf r}_D=(0,0,0.1)~\mu$m, and the two peaks are now less accessible than it in the low pump (linear) regime; for a pump rate around $P_a=40~\mu$eV, {the steady state populations are around $N_a \approx 0.7, N_b \approx 0.5$, and the double peaks merge into a single resonance peak; in contrast, the populations of both QDs are negligible for the pump rate as low as $P_a=0.1~\mu$eV.} 
 It is  interesting to note that  similar physics occurs in an incoherently pumped quantum-dot--cavity system~\cite{YaoPRB81}, though in that case there was no incoherent coupling contribution ($\gamma_{12}=0$), so the two peaks (vacuum Rabi splitting peaks) were at the same height.
In the present case, the doublet feature is entirely due to photon exchange
effects, mimicking the well known vacuum Rabi doublet.

\begin{figure}
\centering\includegraphics[trim=0.4cm .6cm 0.4cm 0cm, clip=true,width=.999\columnwidth]{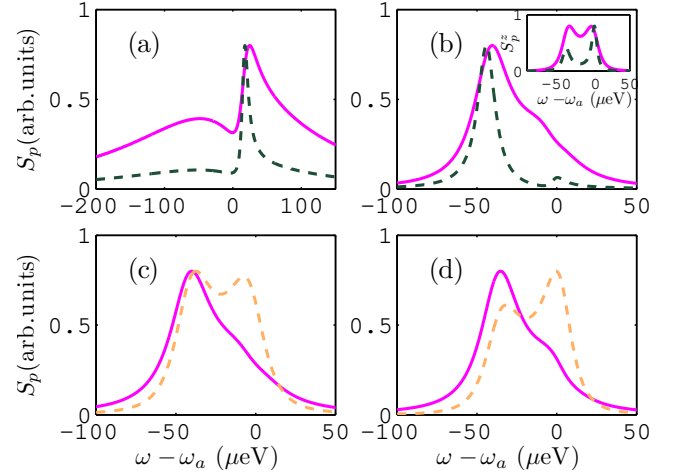}
\caption{(Color online) Incoherent spectra observed at ${\bf r}_D = (0,0.3,0)~\mu$m for $x$-polarized QD$a$ at position ${\bf r}_a = (60,0,0)~$nm and $z$-polarized QDb at ${\bf r}_{b'} = (-45,0,23)~$nm (a) $S_p$ for resonant QDs ($\Delta=0$) with $P_a=0.1~\mu$eV (green dashed) and $P_a=10~\mu$eV (magenta solid); $\gamma'=1~\mu$eV; (b) same as (a) but with $\Delta=\Delta_{\rm off}$; insert shows $S_p^z$ with the same color scheme. (c) $S_p$ (magenta solid) and $S_p^z$ (orange dashed) for resonant QDs ($\Delta=\Delta_{\rm off}$) with $P_a=10~\mu$eV and $\gamma'=5~\mu$eV. (d) $S_p$ (magenta solid) and $S_p^z$ (orange dashed) for off resonant QDs ($\Delta_{ab}=-10~\mu$eV, $\omega_a = \omega_c+\Delta_{\rm off}$) with $P_a=10~\mu$eV and $\gamma'=1~\mu$eV.}
\label{fig:fPL3}
\end{figure} 

As is  discussed above, by using the QMN technique we can efficiently conduct a detailed study of the positional dependence  on the dynamics of the system. By way of an example,  we next study the incoherent spectra of orthogonal QDs at the detector position ${\bf r}_D = (0,0.3,0)~\mu$m,  with an $x$-polarized QD$a$ at ${\bf r}_a$, and using a $z$-polarized QD$b$ at ${\bf r}_{b'}=(-45,0,23)~$nm; for this configuration, note that QD$b$ (polarized in the $z$ direction) obtains an even larger SE enhancement induced by the QNM coupling, $\gamma_b^{\rm qnm}\approx500\,(\gamma_0)$, but a slightly smaller nonradiative contribution of $\gamma_b^{\rm nrq}\approx212\,(\gamma_0)$ when $\Delta_{ab}=0$; the spectra, $S_p$, with $P_a = 0.1/10~\mu$eV, $\gamma'=1~\mu$eV and $\Delta=0$ are shown in Fig.~\ref{fig:fPL3}(a) by the (dark) green dashed/magenta solid lines, respectively. While the spectra, $S_p$, at $\Delta=\Delta_{\rm off}$ are shown in Fig.~\ref{fig:fPL3}(b) by the green dashed line ($P_a=0.1~\mu$eV) and magenta solid line ($P_a=10~\mu$eV); the inset shows $S_p^z$ with the same color scheme; we see that it is now much easier to access the double-peak structure with the polarization dependent spectrum. For a larger pure dephasing rate of $\gamma'=5~\mu$eV, $S_p$ is shown in Fig.~\ref{fig:fPL3}(c) by the magenta solid at $\Delta=\Delta_{\rm off}$, and it shows the double-peak structure is less visible as the pure dephasing rate  increase; the orange dashed line displays $S_p^z$. Finally, we have also studied the effect of detuning between the QDs on the spectra. Figure~\ref{fig:fPL3}(d) shows the spectra with QD-QD detuning $\Delta_{ab}=-10~\mu$eV, $P_a=10~\mu$eV and $\gamma'=1~\mu$eV: the detuning changes both the separation and position of the peaks, and the double peak structure is seen to be robust against detuning as long as it is not too large (with respect to $\delta_{ab}$) as is shown by the magenta solid line ($S_p$). But as the detuning becomes larger and larger the double-peak inevitably begins to disappear. However, this robustness is in general much larger than for narrowband dielectric cavity systems.

In general the splitting of the incoherent spectrum $S_p$ could be effectively controlled by the coherent coupling strength between the QDs ($\delta_{ab}\propto {\rm Re}[{\bf G}_{ab}^{\hat{n}_a\hat{n}_b}(\omega_{a})]$),  which can be achieved by changing both the location and polarization of the QDs (or moving the nanorod); the relative height of the double peak will be changed at the same time since the position dependent behaviour of the plasmon-induced decay rate ($\gamma_n\propto {\rm Im[{\bf G}_{kk}^{\hat{n}_n\hat{n}_n}](\omega_n)}$) and the cross decay rate ($\gamma_{12}\propto {\rm Im}[{\bf G}_{ab}^{\hat{n}_a\hat{n}_b}(\omega_a)]$) will also change.

\section{conclusion}
\label{sec:con}

In summary, we have presented an efficient master equation formalism to include the effect of coupling artificial atoms (QDs) to the dissipative electromagnetic response of a gold nanorod or general shaped metal resonator, which was aided through a QNM expansion of the medium Green function. Using the derived master equation we studied the dynamics of two QDs, and showed that due to the complicated position-dependent polarization characteristic of the LSP, QDs could be effectively coupled together even with orthogonal polarization. Our results first show that the non-Markovian regime can be important for  time scales of the order of the   LSP lifetime, after which  the dynamics of the SE decay is well described by the exponential decay law with decay rate given by the imaginary part of the Green function at the frequency of the QD. 
We have also discussed how our model differs and compares with a simpler JC approach, and the potential limitations of the JC model are highlighted. In certain regimes where a JC model could work, then the required parameters can also be obtained directly from the QNM theory. Using our more general theory, we then presented a selection of examples to study the quantum dynamics of two QDs coupled to a gold nanorod, and discussed the various radiative and nonradiative coupling rates as a function of QD position.
For separate initial states with one of QDs excited, maximum entanglements of greater  than 0.7 could be achieved within a few ps, which also  shows a non-monotonic behavior as a function of distance from the QDs to the nanorod;  we also showed that in order to get the QDs more effectively coupled, the QDs should be detuned away from the resonance of the LSP. With an incoherent pump, Fano resonance features are predicted in emitted spectrum, with a rich polarization dependent behaviour and a double-peak structure that signals strong photon exchange effects between the LSP coupled QDs. Importantly, our theory can quickly treat the coupling dynamics between multiple QDs at various spatial locations over a wide range of frequencies and allows an intuitive understand of the underlying physics of LSP coupling, including a proper decoupling of radiative and nonradiative decay channels.
As shown by  Kewes {\em et al.}~\cite{BuschSPASER}, 
the ability to separate such processes is important to accurately model
emerging nanoplasmonic devices such as SPASERS.

\section*{Acknowledgements}
This work was supported by the Natural Sciences and
Engineering Research Council of Canada and Queen's University.



\begin{thebibliography}{[1]}
\bibitem{comm1}
Note that the real part of the Green function diverges as the two space points approaching each other, which contributes to the homogeneous space  Lamb shift of the QDs; this contribution has been included in the definition of the resonant frequencies $\omega_n$.
\end{thebibliography}
\end{document}